\documentclass[11pt]{cimento_empty}

\usepackage{epsfig}
\usepackage{subfigure}
\usepackage{cite}
\usepackage{amsmath}
\usepackage{mathrsfs}
\usepackage{amssymb}
\usepackage{slashed}

\newcommand{\U}{\mathbf{U}}
\newcommand{\T}{\mathbf{T}}
\newcommand{\V}{\mathbf{V}}
\newcommand{\DL}{\mathcal{D}}
\newcommand{\G}{\mathcal{G}}
\renewcommand{\H}{\mathcal{H}}
\newcommand{\SH}{\mathbf{\Sigma}}
\newcommand{\At}{\tilde{\mathcal{A}}}
\newcommand{\Vt}{\widetilde{\mathbf{V}}}
\newcommand{\Bt}{\widetilde{\mathbf{B}}}
\newcommand{\Wt}{\widetilde{\mathbf{W}}}
\newcommand{\ctil}{\tilde{c}}
\newcommand{\BBd}{B_{\mu\nu}}
\newcommand{\BBu}{B^{\mu\nu}}
\newcommand{\WWd}{W_{\mu\nu}}
\newcommand{\WWu}{W^{\mu\nu}}
\newcommand{\LL}{\mathscr{L}}
\newcommand{\F}{\mathcal{F}}
\renewcommand{\P}{\mathcal{P}}
\renewcommand{\O}{\mathcal{O}}

\DeclareMathOperator{\Tr}{Tr}
\newcommand{\tr}{\Tr}
\DeclareMathOperator{\diag}{diag}
\renewcommand{\to}{\rightarrow}
\newcommand{\de}{\partial}
\newcommand{\nn}{\nonumber}
\newcommand{\hc}{\text{h.c.}}

\newcommand{\e}{\varepsilon}

\title{Higgs physics beyond the SM: the non-linear EFT approach}
\author{I.~Brivio}
\instlist{\inst{Instituto de F\'{\i}sica Te\'orica UAM/CSIC and Departamento de F\'isica Te\'orica,\\
Universidad Aut\'onoma de Madrid, Cantoblanco, 28049 Madrid, Spain\\[1cm]
Talk given at the XXIXth Rencontres de Physique de la Vall\'ee d'Aoste, La Thuile (Italy),\\ 1-7~March~2015 }}

\PACSes{
\PACSit{11.15.Ex}{Spontaneous breaking of gauge symmetries}
\PACSit{12.39.Fe}{Chiral Lagrangians}
\PACSit{12.60.Rc}{Composite models}
\PACSit{14.80.Bn}{Standard-model Higgs bosons}
}

\begin{document}

\maketitle

\begin{abstract}
Depending on whether electroweak physics beyond the Standard Model  is based on a linear or on a non-linear implementation of the electroweak symmetry breaking, a linear or a chiral Effective Lagrangian is more appropriate. 
In this talk, the main low-energy signals that allow to recognize whether the observed “Higgs” scalar is a dynamical (composite) particle or rather an elementary one are presented, in a model-independent way. The patterns of effective couplings produced upon the assumption of specific composite Higgs models are also discussed.
\end{abstract}

\section{Motivation}
The discovery of a light Higgs boson at the LHC~\cite{Aad:2012tfa,Chatrchyan:2012ufa} completes beautifully the Standard Model (SM) description of fundamental interactions. Nonetheless, a number of questions about the theoretical consistency of the theory remain unanswered, and point to the existence of some new physics around or not far from the TeV scale.

In the lack of direct detection of any new particle, a worthy and rather model-independent tool for the study of beyond-Standard Model (BSM) physics is provided by Lorentz and gauge-invariant effective Lagrangians, which respect a given set of symmetries, including the low-energy established ones. In particular, with a light Higgs observed, two main classes of effective Lagrangians are pertinent, depending on how the electroweak (EW) symmetry breaking (EWSB) is assumed to be realized: linearly for elementary Higgs particles or non-linearly for ``dynamical" -typically composite- ones.

In elementary Higgs scenarios, the Higgs particle belongs to an $SU(2)_L$ doublet $\Phi$. If the new physics scale is $\Lambda\gg v$, where $v$ is the EW vacuum expectation value (vev), the corresponding linear effective Lagrangian contains operators weighted by inverse powers of the cutoff scale $\Lambda$: the leading corrections to the SM Lagrangian have then canonical mass dimension $d=6$~\cite{Buchmuller:1985jz,Grzadkowski:2010es,Giudice:2007fh,Low:2009di}. A typical example of possible underlying physics are supersymmetric models.

In dynamical Higgs scenarios, on the other hand, the Higgs particle is a composite field which happens to be a pseudo-goldstone boson (GB) of a global symmetry $\G$, that is broken down spontaneously to a subgroup $\H$ at a scale $\Lambda_s$, corresponding to the masses of the lightest strong resonances. The Higgs mass is protected by the GB shift symmetry, thus avoiding the electroweak hierarchy problem. Explicit realizations include composite Higgs models~\cite{Kaplan:1983fs,Kaplan:1983sm,Banks:1984gj,Georgi:1984ef,Georgi:1984af,Dugan:1984hq,Agashe:2004rs,Contino:2006qr,Gripaios:2009pe,Marzocca:2012zn}.
The most suitable effective Lagrangian for this scenario is a non-linear~\cite{PhysRev.177.2247} or ``chiral" one: a derivative expansion as befits goldstone boson dynamics.

Remarkably, the effective linear and chiral Lagrangians with a light Higgs are in general different. There is not a
 one-to-one correspondence of the leading corrections of both expansions, and one expansion is not the limit of the other unless specific constraints are imposed by hand~\cite{Brivio:2013pma} or follow from  particular dynamics at high energies~\cite{Alonso:2014wta}.
Finding signals which differentiate between these two categories may therefore provide a powerful insight to the origin of the EWSB mechanism.
A few relevant examples of such discriminating signatures will be presented in this talk, together with a brief discussion of the results obtained for specific composite Higgs models ({\it i.e.} specific choices of the coset $\G/\H$).
\section{The effective non-linear Lagrangian for a light Higgs}
The effective low-energy chiral Lagrangian for a light Higgs is entirely
written in terms of the SM fermions and gauge bosons and of the
physical Higgs $h$. 
The SM GBs can be described by a dimensionless unitary matrix~\cite{Appelquist:1980vg,Longhitano:1980iz,Longhitano:1980tm,Feruglio:1992wf,Appelquist:1993ka}:
$
\U(x)=e^{i\sigma_a \pi^a(x)/v}$, $\U(x) \rightarrow L\, \U(x) R^\dagger\,,
$
with $L,R$ denoting respectively the $SU(2)_{L,R}$ global transformations of the scalar potential. 
Higgs  couplings are now (model-dependent) generic functions
\begin{equation}\label{F}
 \F_i(h) = 1+2a_i\frac{h}{v}+b_i\frac{h^2}{v^2}+\dots\,.
\end{equation}
The $SU(2)_L$
structure is absent in them and, as often pointed
out (e.g. refs.~\cite{Grinstein:2007iv,Contino:2010mh}), the resulting effective Lagrangian can
describe many setups including that for a light SM singlet isoscalar.
 
The light dynamical scalar particle $h$ can be described by the phenomenological Lagrangian
\begin{equation}\label{Lchiral}
 \LL_\text{chiral} = \LL_0 + \Delta \LL
\end{equation} 
where the leading order $\LL_0$ is the SM Lagrangian, and $\Delta\LL$ describes any deviation from the SM due to strong-interacting new physics present at scales above the EW one.
The former term reads then
\begin{equation}\label{LL0}
\begin{split}
\LL_0 =& \frac{1}{2} (\de_\mu h)(\de^\mu h) -\frac{1}{4}\WWd^a W^{a\mu\nu}-\frac{1}{4}\BBd\BBu-\frac{1}{4} G^a_{\mu\nu}G^{a\mu\nu}- V (h)\\
 &-\frac{(v+h)^2}{4}\tr[\V_\mu\V^\mu]+ i\bar{Q}\slashed{D}Q+i\bar{L}\slashed{D}L\\
 &-\frac{v+  h}{\sqrt2}\left(\bar{Q}_L\U \mathbf{Y}_Q Q_R+\hc\right)-\frac{v+ h}{\sqrt2}\left(\bar{L}_L\U \mathbf{Y}_L L_R+\hc\right)\,,
\end{split}
\end{equation} 
where \mbox{$\V_\mu\equiv \left(D_\mu\U\right)\U^\dagger$} ($\T\equiv\U\sigma_3\U^\dag$) is the vector (scalar) chiral field transforming in the adjoint of $SU(2)_L$.
The covariant derivative is 
\begin{equation}
D_\mu \U(x) \equiv \de_\mu \U(x) +igW_{\mu}(x)\U(x) - 
                      \frac{ig'}{2} B_\mu(x) \U(x)\sigma_3 \, , 
\end{equation}
with $W_\mu\equiv W_{\mu}^a(x)\sigma_a/2$ and $B_\mu$ denoting the
$SU(2)_L$ and $U(1)_Y$ gauge bosons, respectively. In eq.~\eqref{LL0},
the first line describes the $h$ and gauge boson kinetic terms, and
the effective scalar potential $V(h)$. The second line describes the $W$ and $Z$
masses and their interactions with $h$, as well as the kinetic terms
for GBs and fermions.  Finally, the third line corresponds to the
Yukawa-like interactions written in the fermionic mass eigenstate
basis. A compact notation for the
right-handed fields has been adopted, gathering them into
doublets 
$Q_R$ and $L_R$. $\mathbf{Y}_Q\equiv\diag\left(Y_U,\, Y_D\right)$ and $\mathbf{Y}_L\equiv  \diag\left(Y_\nu,\, Y_L\right)$ are two $6\times6$ block-diagonal matrices containing the
usual Yukawa couplings.

The term $\Delta\LL$ includes all the effective operators with up to four derivatives allowed by Lorentz and gauge symmetries. 
In the bosonic (pure gauge, pure Higgs and gauge-$h$ operators), CP even sector, to which we restrict in this talk\footnote{The bosonic CP odd sector is analyzed in~\cite{Gavela:2014vra}, while a complete basis comprehensive of both bosonic and fermionic operators has been proposed in~\cite{Buchalla:2013rka}. 
}, it can be decomposed as
\begin{align}
\Delta\LL= & c_B\P_B(h) + c_W\P_W(h)+c_G\P_G(h) +c_C \P_C(h) + c_T \P_T(h)+c_H \P_H(h)+\nn\\
   &+c_{\Box H} \P_{\Box H}(h)+\sum_{i=1}^{26} c_i\P_i(h)
\label{DeltaL}
\end{align}
where $c_i$ are model-dependent coefficients, and the operators $\P_i(h)$ are defined by~\cite{Alonso:2012px,Brivio:2013pma}:

\vspace*{-1.2mm}
\noindent\begin{minipage}{.5\textwidth}
\small
\setlength{\jot}{2pt}
\begin{align}
\P_C(h) &= -\frac{v^2}{4}\tr(\V^\mu \V_\mu) \F_{C}(h)\nn\\
\P_T(h) &= \frac{v^2}{4} \tr(\T\V_\mu)\tr(\T\V^\mu) \F_{T}(h)\nn\\
\P_{B}(h) &=-\frac{g'^2}{4}\BBd \BBu \F_B(h) \nn\\
\P_{W}(h) &=-\frac{g^2}{4}\WWd^a W^{a\mu\nu} \F_W(h) \nn\\
\P_G(h) &= -\frac{g_s^2}{4}G_{\mu\nu}^a G^{a\mu\nu}\F_G(h)\nn\\
\P_H(h) &= \frac{1}{2}(\de_\mu h)(\de^\mu h) \F_H(h)\nn\\
\P_{\square H}&=\frac{1}{v^2}(\de_\mu \de^\mu h)^2\F_{\square H}(h) \label{basis}\\
\P_{1}(h) &= gg' \BBd \tr(\T \WWu) \F_{1}(h)\nn\\
\P_{2}(h) &= ig'\BBd \tr(\T[\V^\mu,\V^\nu]) \F_{2}(h)\nn\\
\P_{3}(h)&= ig \tr(\WWd [\V^\mu,\V^\nu]) \F_{3}(h)\nn\\
\P_{4}(h) &= ig' \BBd \tr(\T\V^\mu) \de^\nu\F_{4}(h)\nn\\
\P_{5}(h)&= ig \tr(\WWd\V^\mu) \de^\nu\F_{5}(h)\nn\\
\P_{6}(h) &= (\tr(\V_\mu\V^\mu))^2 \F_{6}(h)\nn\\
\P_{7}(h) &= \tr(\V_\mu\V^\mu) \de_\nu\de^\nu\F_{7}(h)  \nn\\
\P_{8}(h) &= \tr(\V_\mu\V_\nu) \de^\mu\F_{8}(h)\de^\nu\F_{8}'(h)\nn
\end{align}
\end{minipage}
\begin{minipage}{.45\textwidth}
\small
\begin{equation*}
\setlength{\jot}{3pt}
\begin{split}
\P_{9}(h) &= \tr((\DL_\mu\V^\mu)^2) \F_{9}(h)\\
\P_{10}(h) &= \tr(\V_\nu\DL_\mu\V^\mu) \de^\nu\F_{10}(h)\\
\P_{11}(h) &= (\tr(\V_\mu\V_\nu))^2 \F_{11}(h)\\
\P_{12}(h) &= g^2 (\tr(\T\WWd))^2 \F_{12}(h) \\
\P_{13}(h) &= ig \tr(\T\WWd)\tr(\T[\V^\mu,\V^\nu]) \F_{13}(h) \\
\P_{14}(h)  &=g \e^{\mu\nu\rho\lambda} \tr(\T\V_\mu) \tr(\V_\nu W_{\rho\lambda}) \F_{14}(h)\\
\P_{15}(h) &= \tr(\T\DL_\mu\V^\mu)  \tr(\T\DL_\nu\V^\nu) \F_{15}(h)  \\
\P_{16}(h) &= \tr([\T,\V_\nu]\DL_\mu\V^\mu) \tr(\T\V^\nu) \F_{16}(h)  \\
\P_{17}(h) &= ig \tr(\T \WWd) \tr(\T\V^\mu) \de^\nu\F_{17}(h)\\
\P_{18}(h) &= \tr(\T[\V_\mu,\V_\nu])\tr(\T\V^\mu) \de^\nu\F_{18}(h)\\
\P_{19}(h) &= \tr(\T\DL_\mu\V^\mu)\tr(\T\V_\nu) \de^\nu\F_{19}(h)\\
\P_{20}(h) &= \tr(\V_\mu\V^\mu) \de_\nu\F_{20}(h)\de^\nu\F_{20}'(h)   \\
\P_{21}(h) &= (\tr(\T\V_\mu))^2 \de_\nu\F_{21}(h)\de^\nu\F_{21}'(h)  \\
\P_{22}(h) &= \tr(\T\V_\mu)\tr(\T\V_\nu) \de^\mu\F_{22}(h)\de^\nu\F_{22}'(h)\\
\P_{23}(h) &= \tr(\V_\mu\V^\mu) (\tr(\T\V_\nu))^2 \F_{23}(h)\\
\P_{24}(h) &= \tr(\V_\mu\V_\nu)\tr(\T\V^\mu)\tr(\T\V^\nu) \F_{24}(h)\\
\P_{25}(h) &= (\tr(\T\V_\mu))^2 \de_\nu\de^\nu\F_{25}(h)\\
\P_{26}(h) &= (\tr(\T\V_\mu)\tr(\T\V_\nu))^2 \F_{26}(h)
\end{split}
\end{equation*}
\end{minipage}

\section{Phenomenology: constraints on the effective operators}
Some of the chiral effective operators listed in the previous section contribute to quantities that have been well measured in current or past experiments, including triple (TGC) and quartic (QGC) gauge vertices, Higgs couplings to two gauge bosons (HVV) and the EW parameters $S$, $T$, $U$. Therefore, it is possible to set bounds on the coefficients $c_i$ with which they enter the Lagrangian~\eqref{DeltaL} and on some of the parameters $a_i$ appearing inside the associated functions $\F_i(h)$ (see eq.~\eqref{F}).

The results of the global fit are reported in ref.~\cite{Brivio:2013pma}, together with all the details of the renormalization procedure and of the numerical analysis.

\section{Phenomenology: signatures of non-linearity}
 The linear and non-linear EFTs are essentially different from each other in general, and it is of the utmost interest to investigate what are the signatures, at the phenomenological level, that would allow to disentangle which of them describes best Nature.
 
 To this aim, it is illuminating to compare the basis of eq.~\eqref{basis} with a complete set of independent linear operators of dimension six in the bosonic CP even sector, exploiting the correspondence  $  \Phi=(v+h)/\sqrt2 \,\U \left( 0\,\;1\right)^T$.
 Such an analysis is performed in refs.~\cite{Brivio:2013pma,Brivio:2014pfa} for the so-called HISZ linear basis~\cite{Hagiwara:1993ck,Hagiwara:1996kf}. In what follows we implicitly use that linear basis, but limiting the discussion to a few relevant examples.
 
 Two main categories of effects stem from the phenomenological comparison of the effective Lagrangians:
 \begin{enumerate}
  \item Couplings that are predicted to be correlated in the linear parameterization, but that receive contributions from independent operators in the non-linear description.
  This is the case of the non-linear operators $\P_2(h)$ and $\P_4(h)$ (see eq.~\eqref{basis}), that are in correspondence with the unique linear term
  \begin{equation}\label{OB}
   \O_B=\frac{ig'}{2}\BBd D^\mu\Phi^\dag D^\nu\Phi \:\to\: \frac{v^2}{16}\big[\P_2(h)+2\P_4(h)\big]\,,
  \end{equation} 
  where $\P_2(h)$ contributes to the TGCs usually dubbed $\kappa_Z$ and $\kappa_\gamma$, while $\P_4(h)$ introduces the anomalous HVV vertices $A_{\mu\nu}Z^\mu\de^\nu h$ and $Z_{\mu\nu}Z^\mu\de^\nu h$. 
  
  In a linear scenario any departure of one of these couplings from its SM value is expected to be correlated with effects in the other three, since they all receive a contribution from $\O_B$, obviating for the time being all the other possible operators.  Moreover, the relative magnitude of such contributions is fixed by the structure of the covariant derivative $D_\mu\Phi$.
  In the most general non-linear framework, instead, no such correlation is present: deviations in $\{\kappa_Z,\kappa_\gamma\}$ are parameterized in terms of the coefficient $c_2$, while those in the two anomalous HVV vertices are proportional to $c_4$.  This effect is due to the different gauge representation chosen in the two theories for the Higgs field: in the chiral formalism the Higgs particle $h$ is treated as a gauge singlet, independent of the three SM GBs. As a consequence, this framework lacks the strong link between the couplings of the Higgs and those of the longitudinal gauge bosons, which in the linear realization is imposed by the doublet structure of the field $\Phi$.
  A completely analogous analysis holds for another pair of chiral operators, $\P_3(h)$ and $\P_5(h)$, that contribute to the same TGV and HVV vertices as $\P_2(h),\,\P_4(h)$ and correspond to the linear operator
  \begin{equation}
   \O_W=\frac{ig}{2}\WWd^a \,D^\mu\Phi^\dag\sigma^a D^\nu\Phi \:\to\: \frac{v^2}{8}\big[\P_3(h)-2\P_5(h)\big]\,.
  \end{equation} 
 
  In the event of some anomalous observation in either of the couplings mentioned above, the presence or absence of correlations 
would allow for direct testing of the nature of the Higgs boson. 
This is illustrated in fig.~\ref{fig:nl1},
where the results of the combined analysis of the TGV and HVV 
data are projected into
combinations of the coefficients 
\begin{equation}\label{sigmadelta}
\begin{aligned}
\Sigma_B\equiv 4(2c_2+c_4a_4)\,, \qquad\qquad \Sigma_W\equiv 2(2c_3-a_5)\,,\\
\Delta_B\equiv 4(2c_2-c_4a_4)\,, \qquad\qquad \Delta_W\equiv 2(2c_3+a_5)\,,
\end{aligned}
\end{equation} 
defined such that at order $d=6$ in the linear regime
$\Sigma_B=c_B$, $\Sigma_W=c_W$, while $\Delta_B=\Delta_W=0$. 
Notice that in the numerical analysis the complete ensemble of chiral operators is taken into account simultaneously.
In the linear framework (at order $d=6$) there is no equivalent of the variables $\Delta_{B,W}$: therefore the $(0,0)$ coordinate in the right panel of fig.~\ref{fig:nl1} corresponds to the linear regime (at order $d=6$), and any evidence for a departure from this point would represent a smoking gun of a non-linear realization of the EWSB.
Would future data point to a departure from the origin of the first
figure, instead, it would indicate BSM physics irrespective of the linear or
non-linear character of the underlying dynamics.

Analogous (de)correlation effects between couplings with different number of Higgs legs have been discussed in refs.~\cite{Contino2012,Isidori:2013cga}. A more complex example, that involves the six chiral operators $\P_{\Box h},\,\P_{6-10}$ is analyzed in ref.~\cite{Brivio:2014pfa}.
  
\begin{figure}[t]
\centering
\includegraphics[width=.4\textwidth]{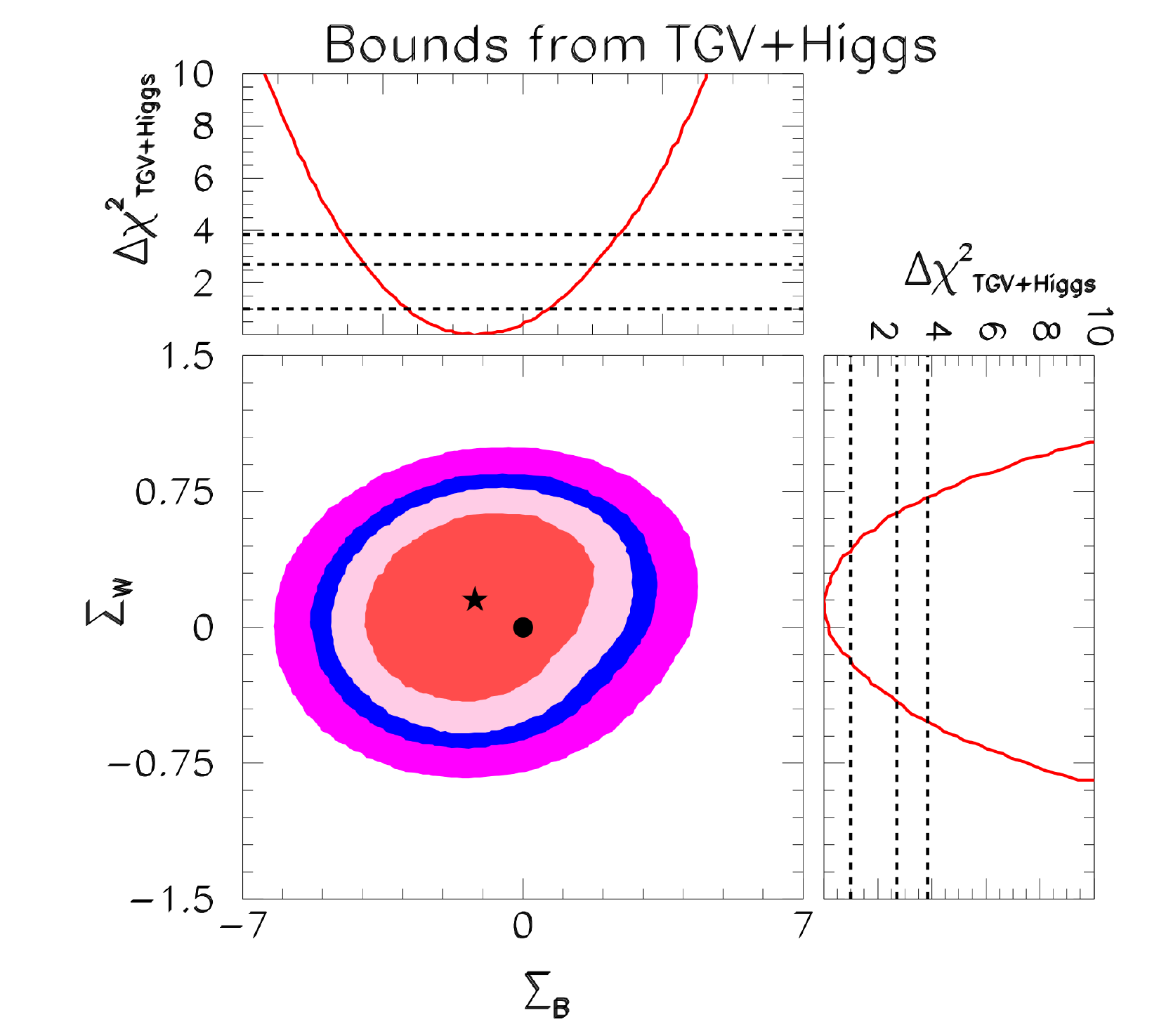}\hfill
\includegraphics[width=.4\textwidth]{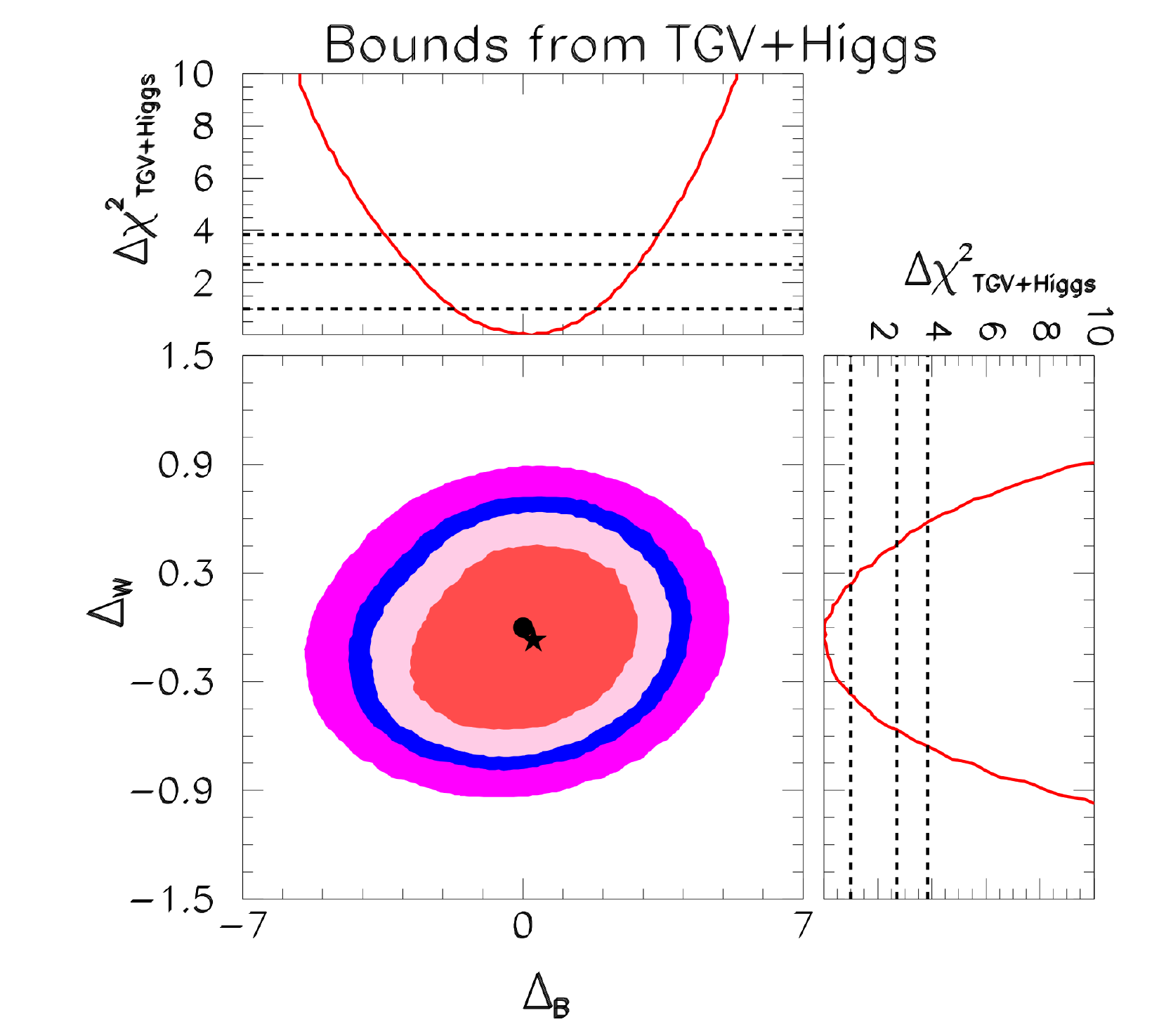}
\caption{\em
{\bf Left}: A BSM sensor irrespective of the type of expansion:
  constraints from  TGV and Higgs data on the combinations
  $\Sigma_B$ and $\Sigma_W$ defined in eq.~\eqref{sigmadelta}, which converge to
  $c_B$ and $c_W$ in the linear $d=6$ limit. 
{\bf Right}: A non-linear versus linear discriminator: constraints on the
combinations $\Delta_B$ and $\Delta_W$. In this case the dot at $(0,0)$ indicates the linear limit.  
Both figures show the allowed regions at 68\%, 90\%, 95\%, and 99\% CL  after marginalization over to the other parameters .
The star corresponds to the best fit point of the analysis. 
}
\label{fig:nl1}
\end{figure}

  \item Couplings that appear only at higher order in the linear expansion, i.e. in linear operators of dimension $d\geq 8$, but are allowed as first-order corrections to the SM (i.e. at the four-derivatives level) in the non-linear description.
  A striking example is that of the operator $\P_{14}(h)$ defined in eq.~\eqref{basis}, whose linear sibling is an operator of dimension 8:
  \begin{equation}
g\e^{\mu\nu\rho\lambda}\WWd^a\,(\Phi^\dag \overleftrightarrow{D}_\rho\Phi)   (\Phi^\dag \sigma^a \overleftrightarrow{D}_\lambda \Phi)\:\to\: v^2\P_{14}(h)\,.
  \end{equation}
The operator $\P_{14}(h)$ contains the anomalous TGC $\e^{\mu\nu\rho\lambda}\de_\mu W^+_\nu W^-_\rho Z_\lambda$, called $g_5^Z$ in the parameterization of~\cite{Hagiwara:1986vm}. Any effect on this coupling, if found to be comparable in size to other leading corrections to the SM, would represent a smoking gun of non-linearity in the EWSB sector. 

  The key to understand how this kind of effect arises is the adimensionality of the $\U(x)$ matrix, which ensures that the GB contributions do not exhibit any scale suppression. This is in contrast with the linear description, where the light $h$ and the three SM GBs are encoded into the scalar doublet $\Phi$, with mass dimension one: in that case any insertion of $\Phi$ pays the price of a suppression factor $1/\Lambda$.

Current limits on the coupling $g_5^Z$ originate from its impact on radiative corrections to $Z$ physics, studied at LEP. However, the LHC has the potential to improve these bounds: the study presented in~\cite{Brivio:2013pma}, based on the kinematical analysis of the process $pp\to W^\pm Z\to\ell'^{\pm}\ell^+\ell^-\slashed{E}_T$, shows that with a luminosity of 300~fb$^{-1}$ at a c.o.m. energy of 14~TeV it is possible to measure $g_5^Z$ at a level of precision comparable to that of the current constraints on dimension 6 linear operators.

 \end{enumerate}

\section{Matching to specific Composite Higgs models}
We now particularize the discussion of the previous sections to the case in which a specific composite Higgs model is assumed, {\it i.e.} in which two specific groups $\G$ and $\H$ are chosen, such that the Higgs particle and the three GBs of the SM arise as the GBs of the spontaneous breaking $\G\to\H$. 

In this kind of scenario, the GBs have a characteristic scale usually denoted by $f$ that satisfies $\Lambda_s \leq 4\pi f$~\cite{Manohar:1983md}. 
At scales above $f$, the model can be described by a high-energy effective Lagrangian $\LL_\text{high}$, manifestly invariant under the complete group $\G$, while the low-energy (below $f$) effects can always be expressed in terms of the effective operators defined in eq.~\eqref{basis}, whose coefficients shall satisfy some model-dependent constraints. 
Studying the connection between $\LL_\text{high}$ and $\LL_\text{chiral}$, it is possible to identify which is, for each composite Higgs realization, the predicted pattern in the coefficients of the low-energy effective chiral operators, an information that may indeed be very valuable when trying to unveil the origin of EWSB. 

The construction of the high-energy effective Lagrangian for a completely generic symmetric\footnote{The condition of a symmetric coset is satisfied in all realistic composite Higgs models. Denoting by $X$ and $T$ any generator of the coset and of the preserved subgroup respectively, it is defined by the schematic constraints $[X,X]\propto T, \quad[T,T]\propto X, \quad[X,T]\propto X$.} coset is reported in detail in ref.~\cite{Alonso:2014wta}, together with the discussion of three specific composite Higgs realizations: the original $SU(5)/SO(5)$ Georgi-Kaplan model~\cite{Georgi:1984af}, the minimal intrinsically custodial-preserving $SO(5)/SO(4)$ model~\cite{Agashe:2004rs} and the minimal intrinsically custodial-breaking $SU(3)/(SU(2)\times U(1))$ model, where by custodial breaking we mean sources of breaking other than those resulting from gauging the SM subgroup.
Here below we summarize the main results.

\subsection{The high-energy effective Lagrangian}
Following the general CCWZ construction \cite{Coleman:1969sm,Callan:1969sn} and under the assumption of a symmetric coset $\G/\H$,  the GBs $\Xi^a(x)$ arising from the spontaneous breaking $\G\to\H$ can be collectively described by the field
$\SH(x)= e^{i\Xi^a(x)X^a/f}$, being $X^a$ the generators of the coset $\G/\H$.
The gauge fields can be introduced as
$ \Wt_\mu\equiv W^a_\mu \,Q^a_L \text{ and } \Bt_\mu \equiv B_\mu \,Q_Y$
where $Q^a_L$ and $Q_Y$  denote the embedding in $\G$ of the $SU(2)_L\times U(1)_Y$ generators. 

The most general Lagrangian describing the interactions of the GBs of a non-linear realization of the symmetric coset $\G/\H$ and of the SM gauge fields can then be written as $\LL_\text{high} = \sum_i \ctil_i\At_i,$ where the sum extends over a complete  basis of bosonic, CP even, operators $\At_i$ with at most four derivatives, constructed out of the structures $\SH, \,\Wt_{\mu\nu},\,\Bt_{\mu\nu}$ and $\Vt_\mu=\left(\mathcal{D}_\mu \SH \right) \SH^{-1}$. Such a basis is reported in eq.~(3.23) of ref.~\cite{Alonso:2014wta}.

Remarkably, the effective Lagrangian at high energy contains at most ten arbitrary coefficients $\ctil_i$, which govern the projection of $\LL_\text{high}$ into $ \LL_\text{chiral}$ defined in eq.~\eqref{Lchiral}. Strong relations are therefore predicted among the many low-energy coefficients of $\LL_\text{chiral}$. Upon specifying the choice of the coset $\G/\H$, the number of free parameters can be further reduced, due to peculiar relations among the generators of $\G$: as an example, the model $SO(5)/SO(4)$ allows for only eight independent high-energy operators.

\subsection{Projection onto the low-energy effective Lagrangian}
For the complete result of the matching, inclusive of the three models considered, we refer the reader to Tables 1 and 2 of ref.~\cite{Alonso:2014wta}. Here we only give the illustrative example of the operator 
\begin{equation}
\At_2= i\,g'\,\Tr\left(\Bt_{\mu \nu}\left[\Vt^\mu,\Vt^\nu\right]\right)\,,
\end{equation}
that for both the $SU(5)/SO(5)$ and $SO(5)/SO(4)$ models decomposes as
$\At_2 = \P_2+2\P_4$, with the dependence on the Higgs field specified by the functions
\begin{equation}\label{sigma_f24}
\begin{split}
 \F_2(h)=\F_4(h)=& \frac{1}{2}\left[1-\left(1-\frac{\xi}{2}\right)\cos\frac{h}{f}+\sqrt\xi\sqrt{1-\frac{\xi}{4}}\sin\frac{h}{f}\right]=\\
 =&\frac{\xi}{4}\left[1+2\sqrt{1-\frac{\xi}{4}}\frac{h}{v}+\left(1-\frac{\xi}{2}\right)\frac{h^2}{v^2}+\O(h^3/v^3)\right]\,,
 \end{split}
\end{equation} 
where $\xi\equiv v^2/f^2$ is a model-dependent parameter that quantifies the degree of non-linearity of the theory: for $\xi\ll1$ the construction converges to the linear realization. As can be easily seen from the second line of eq.~\eqref{sigma_f24}, the dependence on the $h$ field does not respect the linear pattern $(v+h)^2$, which is only recovered in the limit $\xi\to0$. Conversely, the gauge couplings contained in $\P_2$ and $\P_4$ combine with the same relative weight as in the linear $d=6$ description (given by $\O_B$, see eq.~\eqref{OB}). This is a general result of the study performed in~\cite{Alonso:2014wta} that holds for all the three models considered. 

A further tantalising outcome of this analysis is the fact that the $\F(h)$ functions turn out to be the same, up to rescalings of $f$, for all the three models, suggesting that they may be universal to any realistic composite Higgs model. This feature may prove very relevant for the analysis of experimental data, as it provides a precise and almost model-independent signature of composite Higgs  models. More recently, the same subject was explored by ref.~\cite{Low:2014oga}, that drew similar conclusions.

\section*{Acknowledgments}
My work is supported by an ESR contract of the European Union network FP7 ITN INVISIBLES (Marie Curie Actions, PITN-GA-2011-289442). I also acknowledge partial support of the Spanish MINECO’s “Centro de Excelencia Severo Ochoa” Programme under grant SEV-2012-0249.
and I thank the organizers of the La Thuile conference for the kind invitation and for their efforts in organizing this enjoyable meeting.


\end{document}